# Real-Time Stress Measurements in Lithium-ion Battery Negative-electrodes

V.A. Sethuraman,[1] N. Van Winkle,[1] D.P. Abraham,[2] A.F. Bower,[1] P.R. Guduru[1,*]

[1]School of Engineering, Brown University, Providence, Rhode Island 02912, USA
[2]Chemical Sciences and Engineering Division, Argonne National Laboratory, Argonne, Illinois 60439, USA

[*]Corresponding author, Email: Pradeep_Guduru@Brown.edu, Tel: (1) 401 863 3362

Real-time stress evolution in a graphite-based lithium-ion battery negative-electrode during electrolyte wetting and electrochemical cycling is measured through wafer-curvature method. Upon electrolyte addition, the composite electrode rapidly develops compressive stress of the order of 1-2 MPa due to binder swelling; upon continued exposure, the stress continues to evolve towards an apparent plateau. During electrochemical intercalation at a slow rate, the compressive stress increases with the electrode's state-of-charge, reaching a maximum value of 10 – 12 MPa. There appears to be an approximate correlation between the rate of stress rise and the staging behavior of the lithiated graphite. De-intercalation at a slow rate results in a similar linear decrease in electrode stress. Tensile stress of a few MPa develops at the end of deintercalation in the first few cycles, after which the electrode remains under compressive stress only. Although higher peak stresses are seen at high C-rates, the dependence appears to be relatively weak (up to 5C). These measurements reveal, for the first time, the nature of stress evolution in practical lithium-ion-battery electrodes and provide useful data to quantify the driving force for mechanical damage that accrues in composite electrodes upon repeated cycling. While the reported results can serve as reference for calibrating theoretical and computational models to predict electrode stress and damage evolution, the methodology demonstrated can be used to measure stresses and characterize fatigue damage in any composite lithium-ion-battery electrode as well as optimize its microstructure to mitigate stress-related damage mechanisms.

## 1. Introduction

An important aspect of lithium-ion battery research is delineating the mechanisms of the aging processes and modes of degradation that leads to capacity fade [1,2], which is especially important for lithium-ion batteries for automotive applications, where 10–15 years of battery lifetime and a few thousands of charge/discharge cycles are required [3]. Although several active materials are being pursued by researchers worldwide, graphite is still the primary choice for negative-electrodes used in commercial lithium-ion batteries, especially for hybrid and plug-in hybrid electric vehicle (PHEV) applications [4-6]. However, graphitic negative-electrodes suffer from particle cracking and damage resulting in surface-structural disordering upon prolonged cycling [7-11]. These detrimental effects are intensified at high intercalation/de-intercalation rates and elevated temperatures as seen in the Raman spectra of graphite negative-electrodes obtained from aged/cycled lithium-ion cells, which show an increased intensity of the carbon D-band (*ca.* 1350 $cm^{-1}$) with respect to the G-band (*ca.* 1580 $cm^{-1}$) [12-14]. This is particularly true for particles near the separator/negative-electrode interface where the current density is higher than the rest of the composite electrode [7]. Such damage is continuously inflicted on the graphitic crystallites in the negative-electrode upon prolonged intercalation/de-intercalation cycling, and consequently, affects the thickness and composition of the solid-electrolyte-interphase (SEI) layer [15-17].

All mechanical damage in an electrode is driven by the stress and strain fields that are induced during repeated cycling. Hence, in order to quantify the contribution of mechanical damage in a composite electrode to its capacity fade, it is essential to characterize the stress field and its evolution in the electrode. There are numerous theoretical and computational efforts in the literature that predict the stress response of a lithium-ion-battery electrode during the electrochemical intercalation and de-intercalation processes, both at the particle level [18-27] as well as at the electrode level [28-33]. Experimental efforts to directly measure stresses in lithium-ion-battery electrodes are often qualitative in nature [34-38] or limited to idealized planar geometries such as thin films [39-42]. Recently, Qi and Harris carried out *in situ* imaging of a cross section of a composite electrode and measured the deformation field during intercalation/de-intercalation using digital-image correlation [43]. By assuming average material properties, they attempted to calculate the corresponding stress field. There have been no known attempts to directly measure stresses on a practical composite electrode made of a mixture of active material, and a binder during the charge and discharge processes.

The objective of this paper is to demonstrate an experimental technique to directly measure the average stress evolution in the electrode under conditions that closely mimic practical operation of composite lithium-ion-battery electrodes. The stress and strain data were obtained on MAG-10 graphite-based electrode during electrolyte wetting and intercalation/de-intercalation cycles. The stress data obtained in such experiments can be a valuable diagnostic tool in characterizing damage evolution and capacity fade in ageing studies on lithium-battery electrodes. It should be noted that the term composite electrode in this manuscript refers to the mixture of active particles, binder with porosity, which is also known as "coating" in the battery literature.

## 2. Experimental

### *2.1. Electrode preparation*

Detailed information about the fabrication of the composite negative-electrodes and their properties are given in Ref. 44 and in Table 1. Briefly, the negative-electrodes are made of 92% (by weight) MAG-10 graphite particles (Hitachi Powdered Metals Company Ltd., Japan), and 8% PVDF binder (poly-vinylidene fluoride, Kureha KF-1100) with a loading density of 4.9 $mg/cm^2$ (post calendaring). The electrodes were comprised of nominally 35 μm thick, single-side coatings on 18-μm thick Cu current-collector. Figure 1a shows an SEM image of the top surface of the negative-electrode and Figure 1b shows a cross-sectional view of the negative-electrode interior; the images reveal that the graphite particles have a high aspect ratio and are packed mostly parallel to the current collector. The electrodes were cut into circular discs measuring 52 mm in diameter and epoxy bonded (STYCAST 2651) onto Si(111) wafers (single-side polished, 50.8 mm diameter, nominally 250-300 μm thick, with 200 nm thermally grown oxide on all sides). We verified independently that the Si wafer with the thermally grown oxide on it does not participate in the electrochemical reactions; its only role in this experiment is to be an elastic substrate that supports the electrode and facilitates stress measurement. A schematic of the resulting multilayer structure is shown in Figure 2a; a cross-sectional SEM image (post cycling) shown in Figure 2b facilitates accurate measurement of the thickness of each layer.

*2.2. Real-time stress measurements*

Stress evolution in the composite electrode during electrolyte-wetting and the electrochemical-cycling was measured by monitoring the curvature change of the silicon wafer substrate on which the composite electrode is bonded. Substrate curvature was monitored with a multi-beam optical sensor (MOS) wafer-curvature system (k-Space Associates, Dexter, Michigan). The MOS system uses a parallel array of laser beams that get reflected off the sample surface and captured on a camera. The relative change in the spot spacing is related to the wafer curvature through

$$\kappa = \frac{\left(d - d^0\right)}{d^0}\frac{1}{A_m} \qquad 1$$

where $d$ is the distance between two adjacent laser spots on the camera [see figure 1(b) in reference 45], $d_0$ is the initial distance and $A_m$ is the mirror constant, given by $2L/cos(\theta)$; $L$ is the optical path length of the laser beam between the sample and the array and $\theta$ is the incident angle of the laser beam on the sample. The mirror constant $A_m$ is measured by placing a reference mirror of known curvature in the sample plane and measuring the relative change in the spot spacing. Data acquisition rate was 1 Hz for all the experiments.

The Stoney equation is commonly used to determine stresses in a thin film by the wafer-curvature method [45-47]. In the present case, however, the film is no longer "thin" compared to the substrate. Moreover, two additional layers are present, *i.e.*, the epoxy layer and the Cu current collector, the influence of which must be considered in relating the electrode stress to the curvature. The relationship between the average, equi-biaxial stress in the composite-electrode $\sigma^*$, and the substrate-curvature $\kappa$, is obtained as follows. Consider the multilayer structure shown in Figure 2a. Let the equi-biaxial strain at $z = 0$, which is any conveniently chosen plane such as the bottom plane of the substrate, be $\varepsilon_0$. The strain at any other plane along the $z$-direction can be written as,

$$\varepsilon(z) = \begin{cases} \varepsilon_0 - \kappa z, \ 0 \le z \le (h_1 + h_2 + h_3) \\ \varepsilon_0 - \kappa z - \sigma^* / M_4, \ (h_1 + h_2 + h_3) < z \le (h_1 + h_2 + h_3 + h_4) \end{cases} \qquad 2$$

where $h_1$, $h_2$, $h_3$ and $h_4$ are the thickness of the Si substrate, epoxy layer, copper current collector and the composite electrode, respectively, $\kappa$ is the wafer curvature, and $M_4$ is the biaxial modulus of the composite electrode layer, which is defined below. The corresponding stress in each of the layers is given by,

$$
\begin{aligned}
&For\ i\ =\ 1\text{-}3,\ \sigma_i = \frac{E_i}{1-\upsilon_i}\left[\varepsilon_0 - \kappa z\right] = M_i\left[\varepsilon_0 - \kappa z\right] \\
&For\ i\ =\ 4,\ \sigma_4 = \frac{E_4}{1-\upsilon_4}\left[\varepsilon_0 - \kappa z - \sigma^*/M_4\right] = M_4\left[\varepsilon_0 - \kappa z - \sigma^*/M_4\right]
\end{aligned} \quad 3
$$

Where $E_i$, $v_i$ and $M_i$ represent the Young's modulus, Poisson's ratio and the biaxial modulus of the $i^{th}$ layer, respectively. The total potential energy per unit area of the multilayer structure is,

$$
U = \int_0^{h_1} \frac{\sigma_1^2}{M_1} dz + \int_{h_1}^{h_1+h_2} \frac{\sigma_2^2}{M_2} dz + \int_{h_1+h_2}^{h_1+h_2+h_3} \frac{\sigma_3^2}{M_3} dz + \int_{h_1+h_2+h_3}^{h_1+h_2+h_3+h_4} \frac{\sigma_4^2}{M_4} dz \quad 4
$$

The unknowns $\varepsilon_0$ and $\kappa$ are determined by minimizing the potential energy with respect to them, *i.e.*,

$$
\begin{aligned}
&\frac{\partial U}{\partial \varepsilon_0} = 0,\ \text{and} \\
&\frac{\partial U}{\partial \kappa} = 0
\end{aligned} \quad 5
$$

which results in the following expression for $\kappa$

$$
\kappa = \frac{6\sigma^* h_4}{M_1 h_1^2} f\left(h_i, M_i\right) \quad 6
$$

Where $f(h_i,M_i)$ is a function of the thickness and the biaxial modulus of all layers, and is given in the Appendix. The composite-electrode stress, $\sigma^*$ can now be expressed in terms of the substrate curvature as

$$
\sigma^* = \frac{M_1 h_1^2 \kappa}{6 h_4 f\left(h_i, M_i\right)} \quad 7
$$

In using Eq. 7 to analyze the experimental data, it is assumed that $h_4$, the thickness of the composite electrode, remains constant, which is a good approximation for the graphite-based electrodes used in this study.

### *2.3. Electrolyte wetting*

Stress generation in composite electrodes due to binder swelling during electrolyte wetting is an important consideration in evaluating the normal pressure that arises between the electrode roll and battery casing during cell assembly. Binder-swelling induced stresses, in combination with stresses induced during electrochemical cycling are known to cause buckling instabilities in the spiral winding of 18650 type batteries. In addition, electrolyte-wetting-induced stresses make the coin-cell electrodes curl-up during cell assembly, which is often an inconvenience. Furthermore, electrode fabrication conditions are known to influence the extent of curling, which demonstrates the influence of processing parameters on stress generation. Measurement of electrode stress due to wetting by the electrolyte can be a quantitative diagnostic measure to characterize the influence of calendaring conditions on the electrode's electrochemical performance.

To measure electrode stress generated upon electrolyte wetting, woven Celgard C480 separator (thickness = 21 µm, Celgard Inc., Charlotte, North Carolina) was cut into circular discs measuring 55 mm diameter and placed in a cell made of Teflon. The multilayer-electrode structure was then placed in this cell such that the composite electrode faced down (on the separator) while the polished side of the Si wafer faced up. A more detailed description of the experimental arrangement can be found in Ref. 45. Electrolyte (1.2 M $LiPF_6$ in 1:1 by wt. ethylene carbonate:diethyl carbonate) was then added along the periphery of the separator disc with a micro-pipette such that the separator was completely soaked, while the multilayer-electrode structure remained undisturbed. The Si wafer curvature was continuously monitored throughout this process (*i.e.*, before, during, and after the electrolyte addition). Care was taken not to let the electrolyte flood the top surface of the Si wafer, which would otherwise interrupt the wafer-curvature measurements.

*2.4. Electrochemical cycling*

Electrochemical cycling was carried out in a cell described above, with a lithium metal counter and reference electrode at the bottom of the cell. Electrochemical measurements were conducted in ultra-high pure argon atmosphere at 25°C (±1°C) using a Solartron 1470E MultiStat (Solartron Analytical, Oak Ridge, TN). For the first five cycles, lithium was galvanostatically intercalated and de-intercalated from the composite electrode at a current-density of 50 $\mu A/cm^2$ (*ca.* C/36 rate, C = 372 mAh/g denotes the theoretical intercalation capacity of the carbon electrode) between 1.2 and 0.01 V *vs.* $Li/Li^+$. Subsequent intercalation/de-intercalation cycles and stress measurements were performed at progressively increasing rates between C/7 and 5 C. For the higher-rate experiments, the electrode was intercalated using the CCCV protocol (galvanostatic intercalation until the electrode potential reached 10 mV *vs.* $Li/Li^+$, potentiostatic intercalation at 10 mV *vs.* $Li/Li^+$ until the intercalation current decreased to less than 50 $\mu A/cm^2$), and de-intercalation was done galvanostatically.

**3. Results and discussion**

Figures 3(a) and 3(b) show photographs of a graphite-based negative-electrode before and after electrolyte wetting, which results in deformation and curling due to binder swelling-induced stress. Figure 3(c) shows the stress evolution in the electrode upon contact with electrolyte; the stress increases rapidly initially and appears to reach a plateau of *ca.* 1.25 MPa. The stress transient shown in Figure 3(c) is dependent on the tortuosity and pore-size distribution of the composite electrode, mechanical properties of the polymeric binder as well as its volumetric expansion in the electrolyte used. From the measured stress-magnitude, it would be possible to investigate the effect of binder swelling in each electrode layer on the overall pressure build-up in a spirally-wound electrode constrained in a battery casing, a model for which will be presented elsewhere.

Stress evolution during the first three cycles of electrochemical lithiation-delithiation at *ca.* C/36 rate is shown in Figure 4. The average de-intercalation capacity obtained at this rate was 334 mAh/g and the cycling efficiency obtained at the end of the initial formation cycles was 99.3%, both of which are reasonable for a flooded beaker cell with a free-standing electrode. Upon lithiation, the stress becomes compressive and it appears to increase with three distinct slopes, reaching a maximum value in the range of 10 – 12 MPa. Although the correlation is not precise, the different slopes of stress evolution appear to coincide with the different potential plateaus (*i.e.*, graphite staging phenomena), as indicated in Figure 5. The distinct plateaus seen in both the intercalation and de-intercalation potential profiles in Figure 5(a) correspond to stage-4 to stage-1 compounds [49]. The potential plateau at *ca.* 200 mV corresponds to the transition to a stage-3 graphite-intercalation compound (GIC) from a dilute stage-1 *via* stage-4, the potential plateau at *ca.* 110 mV *vs.* $Li/Li^+$ corresponds to a transition from stage-3 GIC to an ordered stage-2 GIC, and the potential plateau at *ca.* 72 mV *vs.* $Li/Li^+$ corresponds to a transition from an ordered stage-2 GIC to a stage-1 GIC [49,50]. Although the mechanisms responsible for the characteristic shapes of stress evolution at low intercalation/de-intercalation rates need to be investigated further, it may be related either to the rate of increase in elastic modulus of graphite with lithiation [51] or to the stage-dependent volume expansion of the graphite-intercalation electrode [11]. During delithiation, the stress initially decreases rapidly, followed by a gradual relaxation as more lithium is extracted. The initial rapid drop appears to be due to elastic unloading of the active particles, which expand, get compressed against each other and deform during lithiation. Rate of stress relaxation appears to correlate with staging during de-lithiation as well (Fig. 5). At the end of delithiation, the stress becomes slightly tensile in the first few cycles (Figure 4), indicating inelastic deformation of the particles and/or the binder during lithiation.

Note that the stress measurements reported here are average values over the composite-electrode thickness and microstructure. The point-wise stress values would be very non-uniform and would vary significantly depending on the local geometry, as evidenced in the electrochemical-strain-microscopy studies on various active materials at the sub-micron scale reported by Kalinin and coworkers [52-55]. A detailed 3D numerical simulation would be needed to construct the stress field at the microstructural level. The stress values measured and reported here, for the first time, quantify the driving force for mechanical damage that accumulates in the electrode upon repeated cycling. Moreover, the stress evolution is clearly a function of the electrode composition, properties of each constituent material and the microstructure. Since mechanical degradation is driven by the stress field, it is reasonable to expect that there is an optimal combination of material properties and microstructure which maximizes battery life. The experimental effort presented here, when combined with detailed 3-D modeling effort, can quantify the influence of material properties and microstructure on the battery life.

In order to highlight the electrode potentials at which large stress changes occur, Figure 6 presents the stress *vs.* potential plot for the data shown in the preceding figure. Significant changes in the rate of stress evolution are seen to occur at potentials that coincide with the onset of staging-induced plateau potentials seen in Figure 5(a). As noted above, such changes may be indicative of concentration-dependent changes in the mechanical properties of lithiated graphite. For instance, using density functional theory, Qi *et al.* [51] have reported that the Young's modulus of Graphite increases threefold upon lithium intercalation to $LiC_6$. Further investigations of the incremental changes in the mechanical properties of graphite and graphite-like intercalation compounds during lithium intercalation are required to fully understand this phenomenon. The stress-potential relation shown in Figure 6 could potentially serve as a useful tool to identify potential ranges that contribute to large stress increases; it might be possible to use such information in mitigating mechanical damage.

Figure 7 shows stress evolution at a higher intercalation/de-intercalation rate (*ca.* 1C), showing a slight increase in peak compressive stress (12-13 MPa) compared to the initial slow rate cycles (10-12 MPa). During a sequence of about 10 intercalation/de-intercalation cycles conducted at this rate, stress response showed minimal cycle-to-cycle variation. Note that the initial rapid stress drop is approximately 65% of the total stress amplitude, which happens during the initial 15% of delithiation. Such a response presents a potential opportunity to conduct accelerated fatigue studies on these electrodes by delithiating and relithiating only about 15% of the capacity, while still subjecting the electrode to 65% of the peak stress amplitude; results from such a study will be presented elsewhere.

Surface and cross sectional images of the negative-electrode particles obtained by SEM/FIB sectioning did not show any obvious evidence of particle cracking due to the limited number of cycles that the electrode was subjected to; however, microfracturing of the particle edges at high cycling rates cannot be ruled out. Figure 8 shows the potential-, current density and stress-hysteresis plots for a cycle each at three very different C-rates. The higher polarization losses at 3C- and 5C-rate data could be due to the polarization of lithium-metal counter electrode as well as ohmic losses. For higher-than-C de-intercalation rates, the stress in the electrode monotonically decreases with lithium concentration and does not show the correspondence with staging behavior seen at slower rates (*e.g.,* Figure 5). Note that the area of the hysteresis loop in the potential-capacity plot is the total electrical-energy loss in the cycle; and the area of the stress-capacity plot represents the energy dissipation due to mechanical deformation of the electrode, *i.e.*, loss due to plastic deformation of the binder, particles sliding at the interfaces, *etc.* A more detailed study is necessary to correlate the microscopic damage modes with the stress history during a large number of cycles.

Furthermore, the stress measurements reported here for various intercalation/de-intercalation rates may serve as a reference for models predicting stresses in similar graphite-based composite electrodes. Wide discrepancies in predicted values of stress distribution across the thickness of the electrode currently exist among various models [18-31]. For example, Christensen recently computed stresses in a porous composite electrode similar to the one shown in Figures 1 and 2, and reported that the stresses in the negative electrode can vary between 5 MPa and 100 MPa during high-rate discharge for various levels of non-ideality [29]. On the other hand, Renganathan *et al.* predicted uniform stresses in the 5-12 MPa range across the thickness of the negative electrode with very little dependence on the rate of discharge between C/2 and 3C [30]. Modeling stresses in a lithium-ion battery electrode to a reasonable degree of accuracy is a difficult and challenging exercise because of the lack of experimentally-measured property data for various constituents of the composite electrode as well as its complex geometry. However, modeling tools are very important in understanding the influence of microstructural and processing parameters on electrode performance. The stress-measurement methodology developed and the data reported in this work for graphite-based negative electrodes is expected to serve as baseline reference to validate the accuracy of simulation efforts.

## 4. Conclusions

Real-time stress measurements on practical composite lithium-ion battery negative electrodes are reported. Upon electrolyte addition, the composite electrode rapidly develops compressive stress of the order of 1-2 MPa due to binder swelling, which evolves towards a plateau. During electrochemical intercalation, the compressive stress increases, reaching a maximum value of 10 – 12 MPa. There appears to be an approximate correlation between the rate of stress rise and the staging behavior of the lithiated graphite. Tensile stress of a few MPa develops at the end of de-intercalation in the first few cycles, after which the electrode is under compressive stress only. The measurements reported in this study reveal the nature of stress evolution in practical lithium-battery electrodes and provide useful data to quantify the driving force for mechanical damage in them. The stress-measurement methodology demonstrated here has the potential to serve as a diagnostic tool to assess the long-term mechanical performance and capacity fading of any practical electrodes after systematically correlating the stress history with microstructural damage evolution due to repeated cycling. Further, the reported technique is expected to be useful in measuring stress evolution in composite electrodes made of active materials that exhibit large volume expansion upon lithiation (*i.e.,* Al, Si, Sn, *etc.*), where mechanical damage is of greater concern than in graphite based electrodes. In combination with complementary modeling efforts, the experimental technique presented here is expected to quantify the influence of material properties, microstructural parameters and processing variables on the mechanical degradation modes, which can lead to systematic optimization of these quantities for improved battery life and performance.

**5. Acknowledgements**

The authors gratefully acknowledge the funding from the DOE EPSCoR Implementation award (grant # DE-SC0007074), Argonne National Laboratory (contract # 1F-31882), and NASA EPSCoR (Grant # NNX10AN03A).

## Appendix

From equations 3 and 4 we get,

$$\begin{aligned} U = {} & M_1 \int_0^{h_1} \left(\varepsilon_0 - \kappa z\right)^2 dz + M_2 \int_{h_1}^{h_1+h_2} \left(\varepsilon_0 - \kappa z\right)^2 dz \\ & + M_3 \int_{h_1+h_2}^{h_1+h_2+h_3} \left(\varepsilon_0 - \kappa z\right)^2 dz + M_4 \int_{h_1+h_2+h_3}^{h_1+h_2+h_3+h_4} \left(\varepsilon_0 - \kappa z - \varepsilon^*\right)^2 dz \end{aligned} \qquad 8$$

Assuming equilibrium conditions, and applying the principle of minimum potential energy as described above, the curvature κ can be expressed as,

$$\kappa = \frac{6\sigma^* h_4}{M_1 h_1^2} f\left(h_i, M_i\right) \qquad 9$$

$$f(h_i, M_i) = \frac{\left[1 + \frac{2h_2}{h_1} + \frac{2h_3}{h_1} + \frac{h_4}{h_1} + \frac{h_2^2 M_2}{h_1^2 M_1} + \frac{2h_2 h_3 M_2}{h_1^2 M_1} + \frac{h_2 h_4 M_2}{h_1^2 M_1} + \frac{h_3^2 M_3}{h_1^2 M_1} + \frac{h_3 h_4 M_3}{h_1^2 M_1}\right]}{\left[\begin{aligned} & 1 + \frac{4h_2 M_2}{h_1 M_1} + \frac{6h_2^2 M_2}{h_1^2 M_1} + \frac{4h_2^3 M_2}{h_1^3 M_1} + \frac{h_2^4 M_2^2}{h_1^4 M_1^2} + \frac{4h_3 M_3}{h_1 M_1} + \frac{12h_2 h_3 M_3}{h_1^2 M_1} + \frac{12h_2^2 h_3 M_3}{h_1^3 M_1} + \frac{6h_3^2 M_3}{h_1^2 M_1} + \frac{12h_2 h_3^2 M_3}{h_1^3 M_1} + \\ & \frac{4h_3^3 M_3}{h_1^3 M_1} + \frac{4h_2^3 h_3 M_2 M_3}{h_1^4 M_1^2} + \frac{6h_2^2 h_3^2 M_2 M_3}{h_1^4 M_1^2} + \frac{4h_2 h_3^3 M_2 M_3}{h_1^4 M_1^2} + \frac{h_3^4 M_3^2}{h_1^4 M_1^2} + \frac{4h_4 M_4}{h_1 M_1} + \\ & \frac{12h_2 h_4 M_4}{h_1^2 M_1} + \frac{12h_2^2 h_4 M_4}{h_1^3 M_1} + \frac{12h_3 h_4 M_4}{h_1^2 M_1} + \frac{24h_2 h_3 h_4 M_4}{h_1^3 M_1} + \frac{12h_3^2 h_4 M_4}{h_1^3 M_1} + \frac{6h_4^2 M_4}{h_1^2 M_1} + \frac{12h_2 h_4^2 M_4}{h_1^3 M_1} + \\ & \frac{12h_3 h_4^2 M_4}{h_1^3 M_1} + \frac{4h_4^3 M_4}{h_1^3 M_1} + \frac{4h_2^3 h_4 M_2 M_4}{h_1^4 M_1^2} + \frac{12h_2^2 h_3 h_4 M_2 M_4}{h_1^4 M_1^2} + \frac{12h_2 h_3^2 h_4 M_2 M_4}{h_1^4 M_1^2} + \frac{6h_2^2 h_4^2 M_2 M_4}{h_1^4 M_1^2} + \\ & \frac{12h_2 h_3 h_4^2 M_2 M_4}{h_1^4 M_1^2} + \frac{4h_2 h_4^3 M_2 M_4}{h_1^4 M_1^2} + \frac{4h_3^2 h_4 M_3 M_4}{h_1^4 M_1^2} + \frac{6h_3^2 h_4^2 M_3 M_4}{h_1^4 M_1^2} + \frac{4h_3 h_4^3 M_3 M_4}{h_1^4 M_1^2} + \frac{h_4^4 M_4^2}{h_1^4 M_1^2} \end{aligned}\right]} \qquad 10$$

**Table 1: Comprehensive set of properties of individual constituents of the composite negative-electrode.**

| Description | Value | Comments |
|---|---|---|
| **I. Current collector: Copper** | | |
| *Thickness* | 18 µm | |
| *Weight* | 160.56 $g/m^2$ | |
| *Conductivity* | $6 \times 10^5\ \Omega^{-1}cm^{-1}$ | |
| **II. Active material: MAG-10, Graphitic carbon (Hitachi)** | | |
| *Average particle size* | 10.1 µm | Based on laser-diffraction measurement |
| *Active area / Electrode volume* | 3696 $cm^{-1}$ | Based on laser-diffraction measurement |
| *Material capacity (theoretical)* | 372 mAh/g | |
| *Material capacity (actual)* | 334 mAh/g | Based on C/36 rate (Measured*) |
| *Material capacity (actual)* | 322 mAh/g | Based on C/7 rate (Measured*) |
| *Electrode capacity (theoretical)* | 1.82 $mAh/cm^2$ | |
| *Electrode capacity (actual)* | 1.64 $mAh/cm^2$ | Based on C/36 rate (Measured*) |
| *Electrode capacity (actual)* | 1.57 $mAh/cm^2$ | Based on C/7 rate (Measured*) |
| *Active material weight fraction* | 92% | |
| *Active material volume fraction* | 62.3% | |
| *Material density* | 2.25 $g/cm^3$ | |
| *Loading density* | 49 $g/m^2$ | |
| **III. Binder: PVDF (Kureha #C)** | | |
| *Material density* | 1.78 $g/cm^3$ | |
| *Loading density* | 4.261 $g/m^2$ | |
| *Binder weight fraction* | 8% | |
| *Binder volume fraction* | 6.8% | |
| *Electrode porosity* | 30.9% | |

* - measured in a flooded beaker-cell setup used in this study.

**Table 2: Parameters used for the stress calculations presented in this study.**

| Parameter | Definition | Value | Comments |
|---|---|---|---|
| $d_f$ | Electrode diameter (Si wafer diameter) | 5.08 cm | Measured |
| $E_1$ | Young's modulus of Si (111) wafer | 169 GPa | Ref. 56 |
| $E_2$ | Young's modulus of epoxy layer | 4.3 GPa | Ref. 57-59 |
| $E_3$ | Young's modulus of Cu | 117 GPa | Ref. 60 |
| $E_{Gr}$ | Young's modulus of graphite | 10 GPa | Ref. 61,62 |
| $E_b$ | Young's modulus of PVDF | 2 GPa | Ref. 63 |
| $E_4$ | Young's modulus of the composite layer | 6.36 GPa | Calculated |
| $h_1$ | Substrate thickness | 450 μm | Measured |
| $h_2$ | Epoxy-layer thickness | 60 μm | Measured |
| $h_3$ | Current-collector thickness | 18 μm | Measured |
| $h_4$ | Active-material thickness | 47 μm | Measured |
| $M_1$ | Biaxial modulus of Si (111) wafer | 228.3 GPa | Ref. 56 |
| $M_2$ | Biaxial modulus of epoxy layer | 7.8 GPa | Calculated |
| $M_3$ | Biaxial modulus of Cu | 181.8 GPa | Calculated |
| $M_4$ | Biaxial modulus of the composite layer | 8.03 GPa | Calculated |
| $v_{Gr}$ | Volume fraction of graphite | 0.622 | Measured |
| $v_b$ | Volume fraction of binder | 0.068 | Measured |
| $v_p$ | Volume fraction of pores | 0.309 | Measured |
| $2L/cos(\theta)$ | Mirror constant | 2.8 m | Measured |
| $\upsilon_1$ | Poisson's ratio of Si (111) substrate | 0.26 | Ref. 56 |
| $\upsilon_2$ | Poisson's ratio of epoxy resin | 0.36 | Ref. 58 |
| $\upsilon_3$ | Poisson's ratio of Cu | 0.347 | Ref. 64 |
| $\upsilon_{Gr}$ | Poisson's ratio of graphite | 0.3 | Ref. 65 |
| $\upsilon_b$ | Poisson's ratio of PVDF | 0.38 | Ref. 66 |
| $\upsilon_4$ | Poisson's ratio of the composite layer | 0.2086 | Calculated |

## Figure-captions

Figure 1: (a) SEM image of an as-received MAG-10 composite negative-electrode, and (b) SEM image of the steepest wall of a staircase trench dug by a dual-beam focused-ion-beam (FIB) system. The angle between the electron beam and the plane of this trench-wall is 52°. The vertical scale bar measuring 6.81 μm shown on the image is corrected for this viewing angle. The bright layer on the top of the trench is a thin Pt layer (*ca.* 1 μm) deposited prior to the FIB process in order to preserve the top surface of the microstructure from getting damaged by the ion-beam.

Figure 2: (a) Schematic illustration of the composite electrode bonded to the Si-wafer substrate with an epoxy layer. Note that the layer thicknesses are not drawn to scale. (b) Cross-sectional SEM image of the composite-wafer structure obtained after the electrochemical-cycling experiments.

Figure 3: Photographs of a composite-electrode piece (a) before, and (b) after 2 hours of immersion in the electrolyte. In the scale bar, each division represents 1/32 inch (0.794 mm). Compressive stresses in the electrode cause the composite structure to bend. (c) Real-time stress evolution data obtained using the wafer-curvature method during an electrolyte wetting experiment.

Figure 4: Transient (a) Potential, and (b) stress response of the composite negative-electrode measured simultaneously during the first three intercalation/de-intercalation cycles. The composite electrode was cycled galvanostatically at 50 $\mu A/cm^2$ (*ca.* C/36 rate).

Figure 5: (a) Potential, and (b) stress data corresponding to the initial intercalation/deintercalation of the composite electrode at *ca.* C/36 rate. The arrows indicate the correspondence between slope changes in the stress plot and the staging-induced potential plateaus, which are labeled for clarity.

Figure 6: Stress *vs.* voltage relation during the first intercalation-deintercalation cycle, highlighting the relation between the staging behavior and stress evolution; the potential ranges in which stress changes rapidly are readily identified. The dashed and dotted lines represent staging-onset potentials during intercalation and deintercalation, respectively. The labels (a-f) indicate onset potentials for stage transitions (*e.g.,* '*a*' represents the onset potential for transition from stage-4 to stage-3).

Figure 7: Representative (a) potential, (b) current, and (c) stress data during a long sequence of intercalation/de-intercalation cycles at *ca.* C rate. The composite electrode was intercalated using a CCCV protocol (*i.e.,* constant current of 1.25 $mA/cm^2$ until 10 mV *vs.* $Li/Li^+$ followed by constant potential at 10 mV *vs.* $Li/Li^+$ until the current is less than 50 $\mu A/cm^2$) and deintercalated galvanostatically at 1.25 $mA/cm^2$ (*ca.* C rate).

Figure 8: A set of potential, current and stress *vs.* capacity plots corresponding to high rate experiments.

**Figure 1**

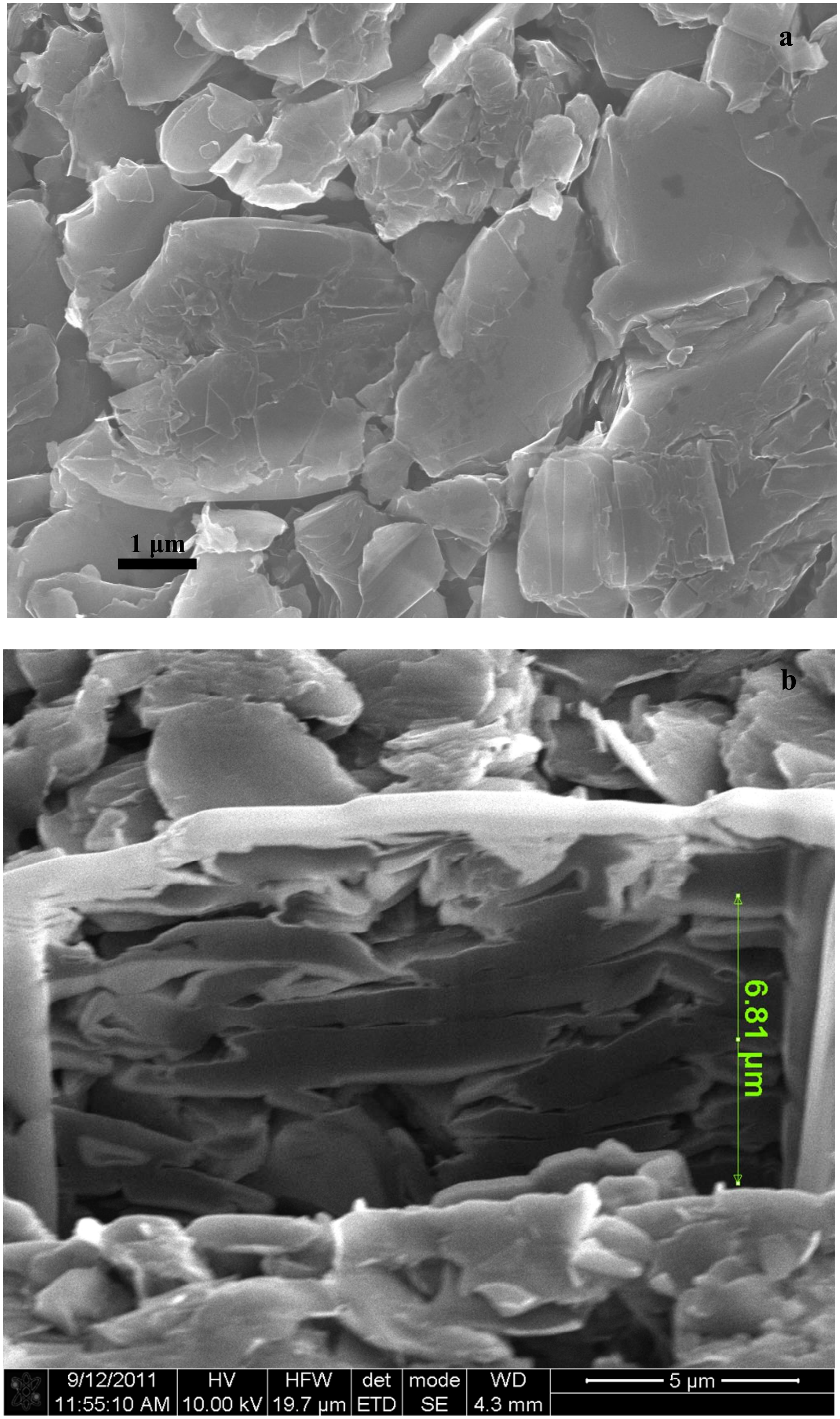

# Figure 2

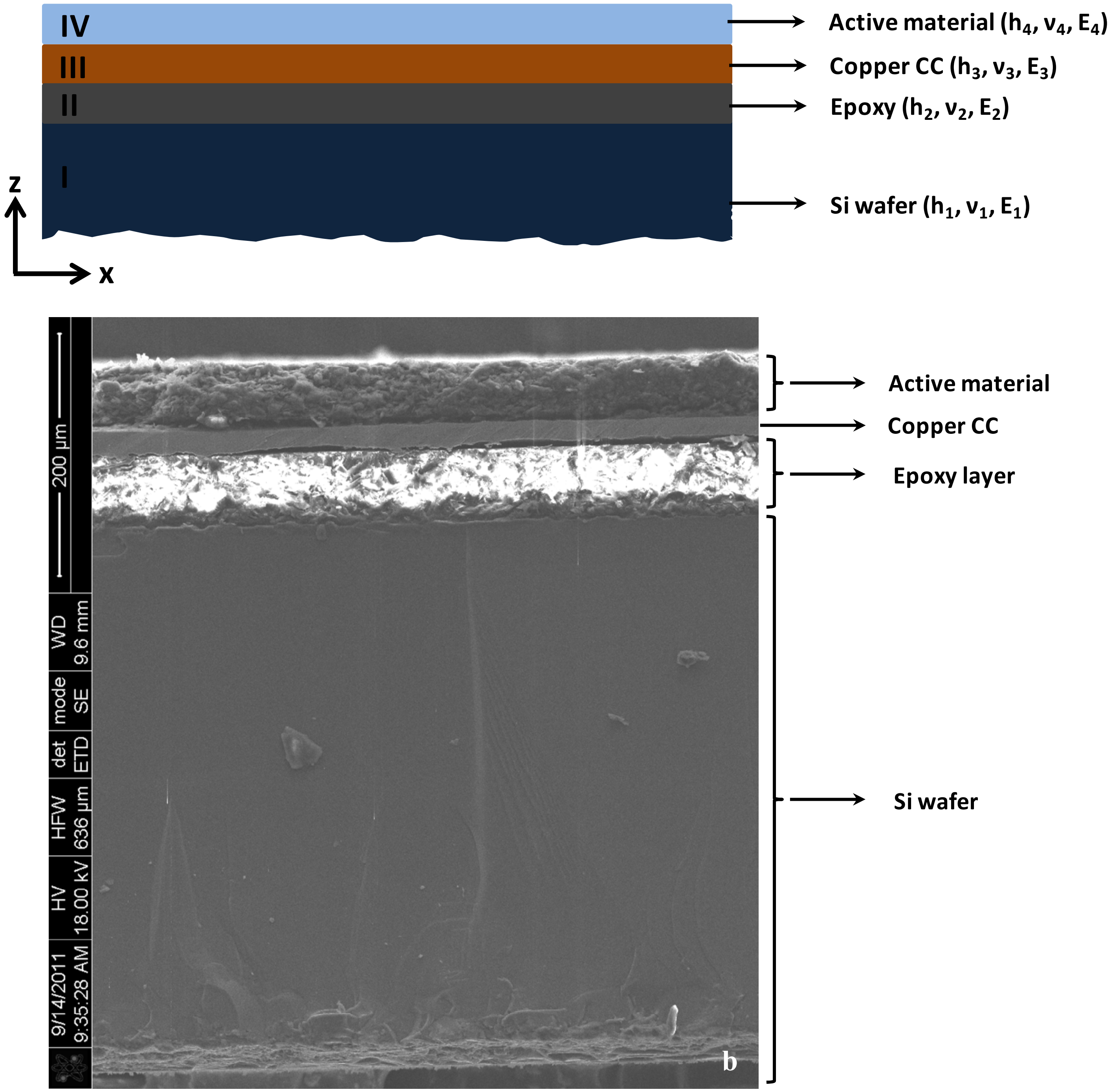

**Figure 3**

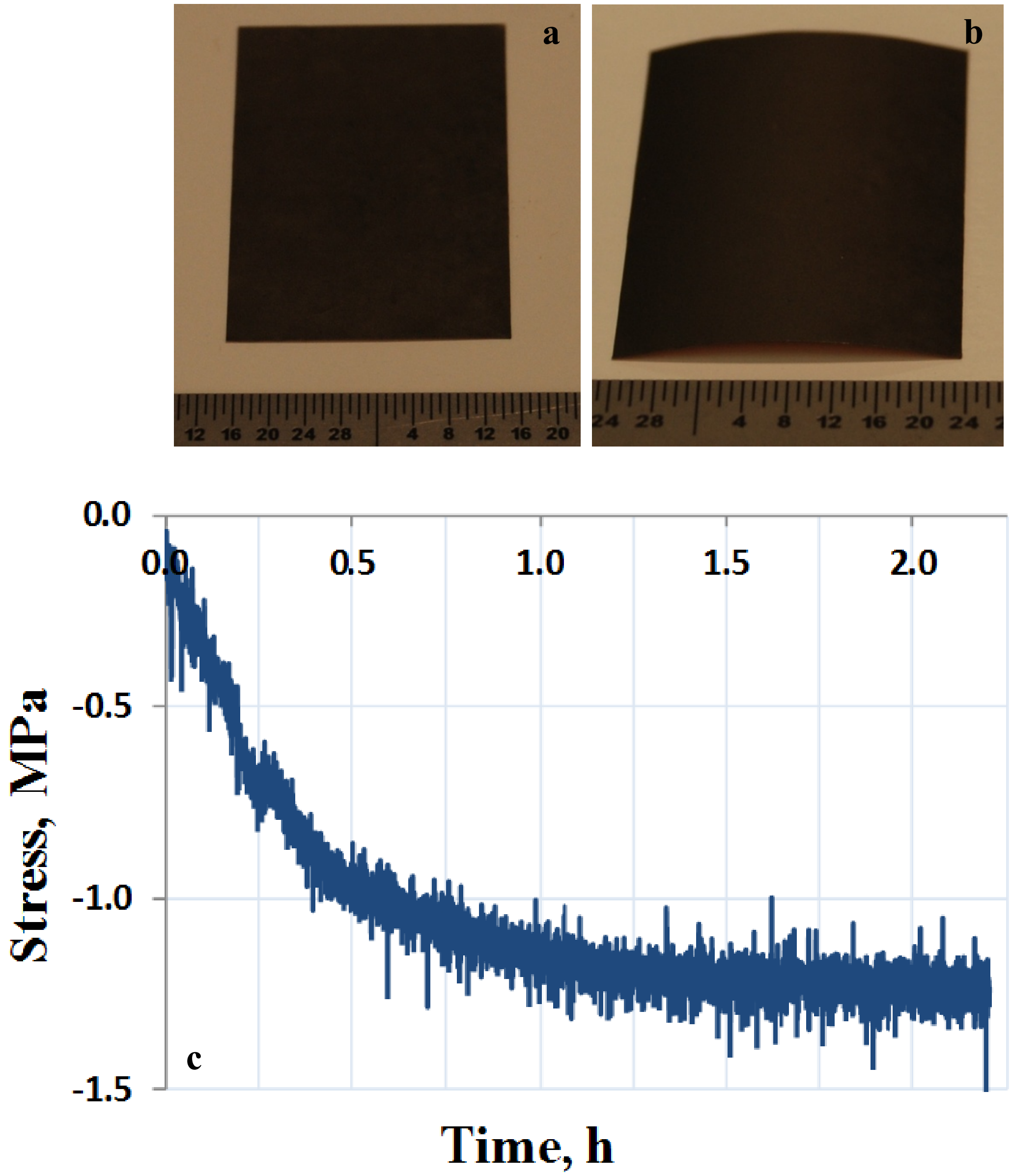
a
12 16 20 24 28 4 8 12 16 20
b
24 28 4 8 12 16 20 24
0.0
-0.5
-1.0
-1.5
0.0
0.5
1.0
1.5
2.0
Stress, MPa
Time, h
c

**Figure 4**

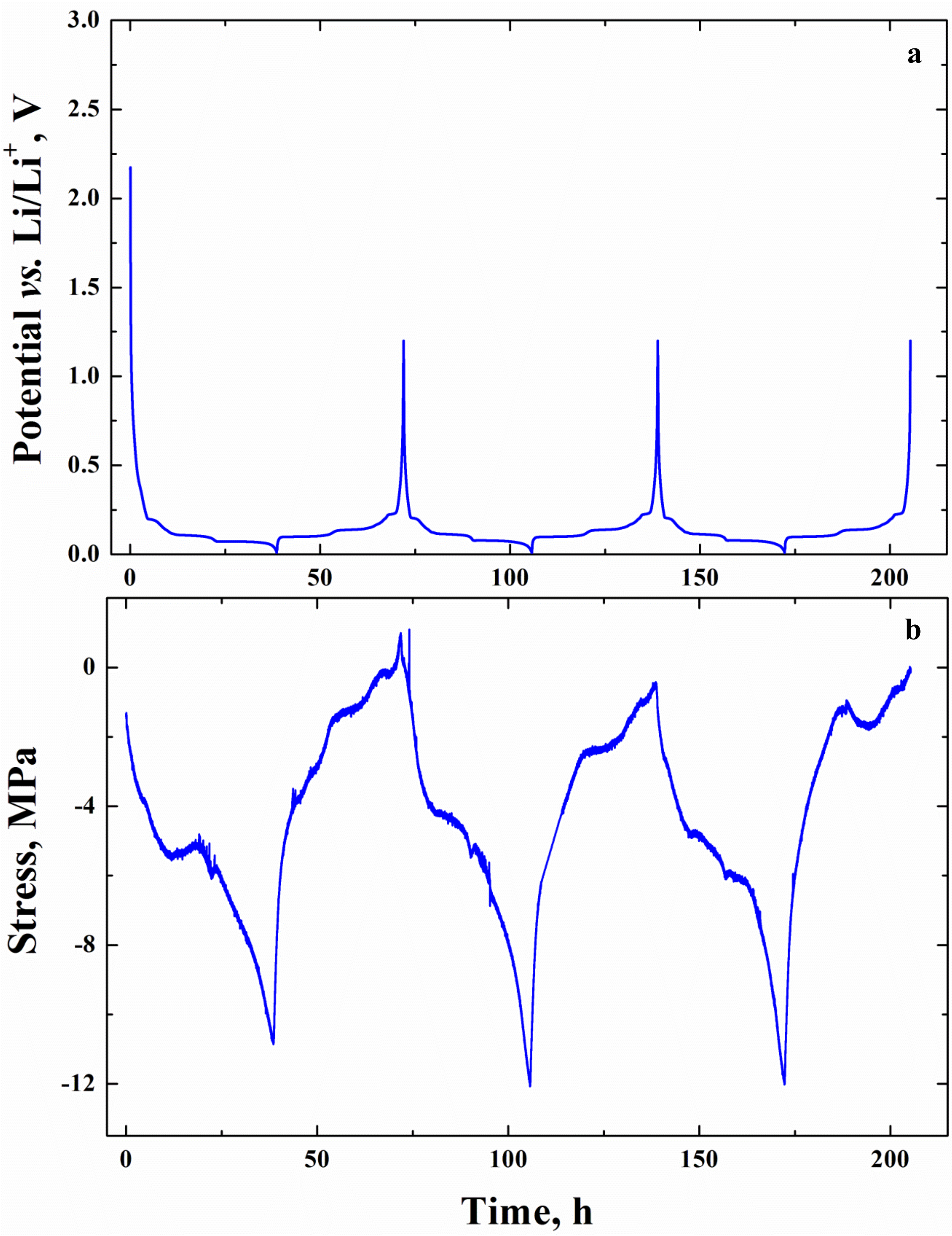

**Figure 5**

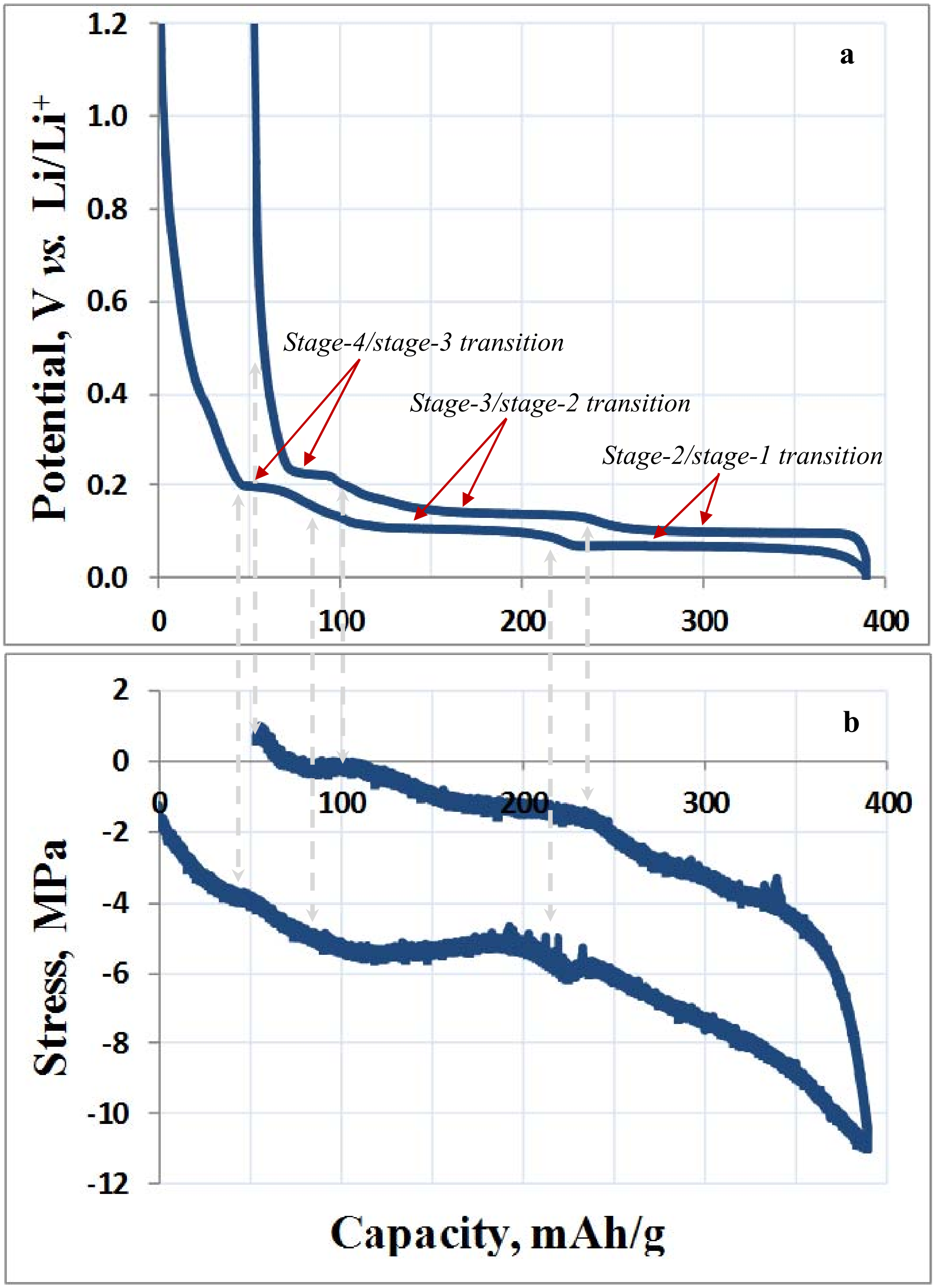

**Figure 6**

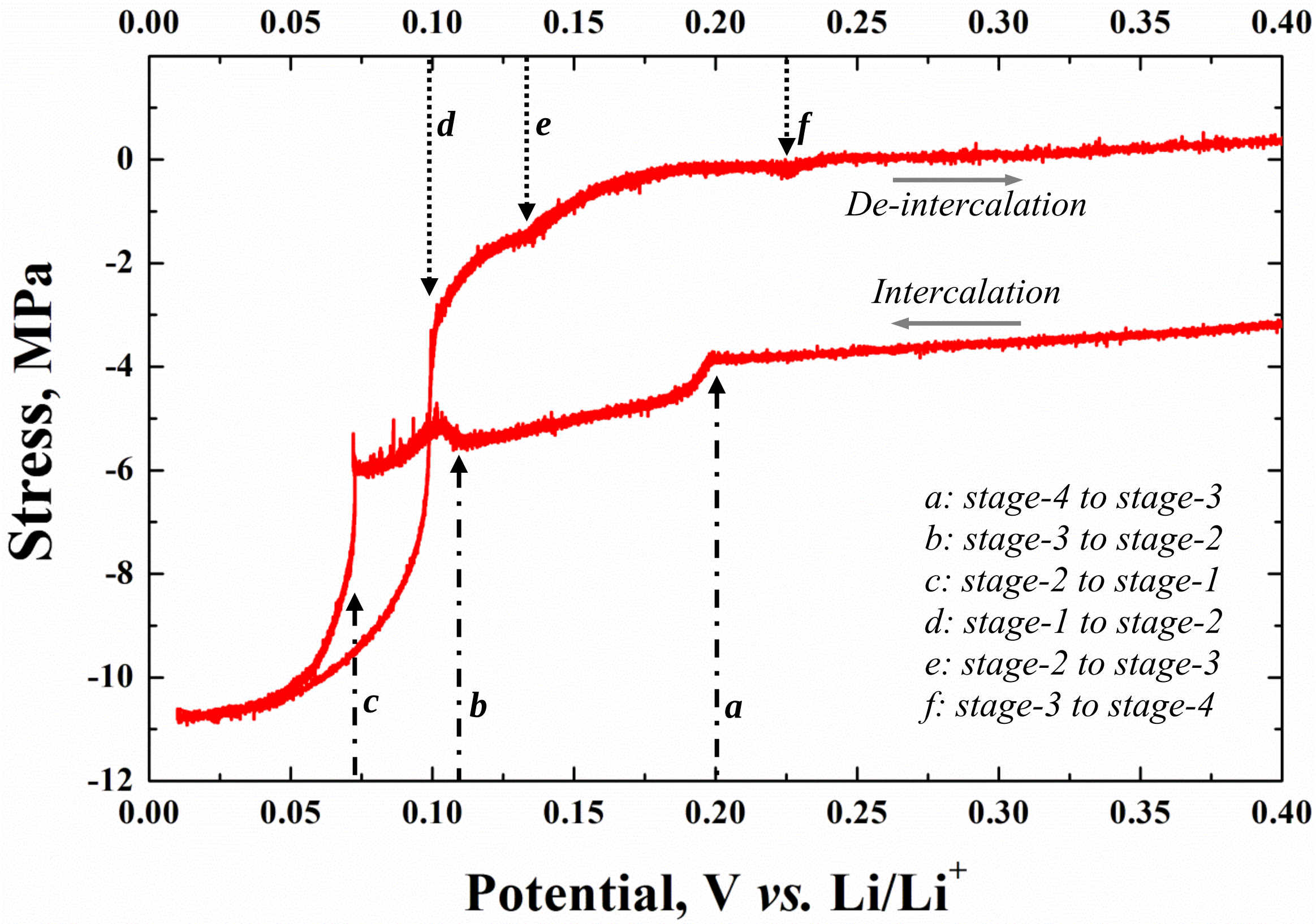

0.00
0.05
0.10
0.15
0.20
0.25
0.30
0.35
0.40
d
e
f
0
-2
-4
-6
-8
-10
-12
De-intercalation
Intercalation
Stress, MPa
a: stage-4 to stage-3
b: stage-3 to stage-2
c: stage-2 to stage-1
d: stage-1 to stage-2
e: stage-2 to stage-3
f: stage-3 to stage-4
c
b
a
Potential, V vs. Li/Li+

# Figure 7

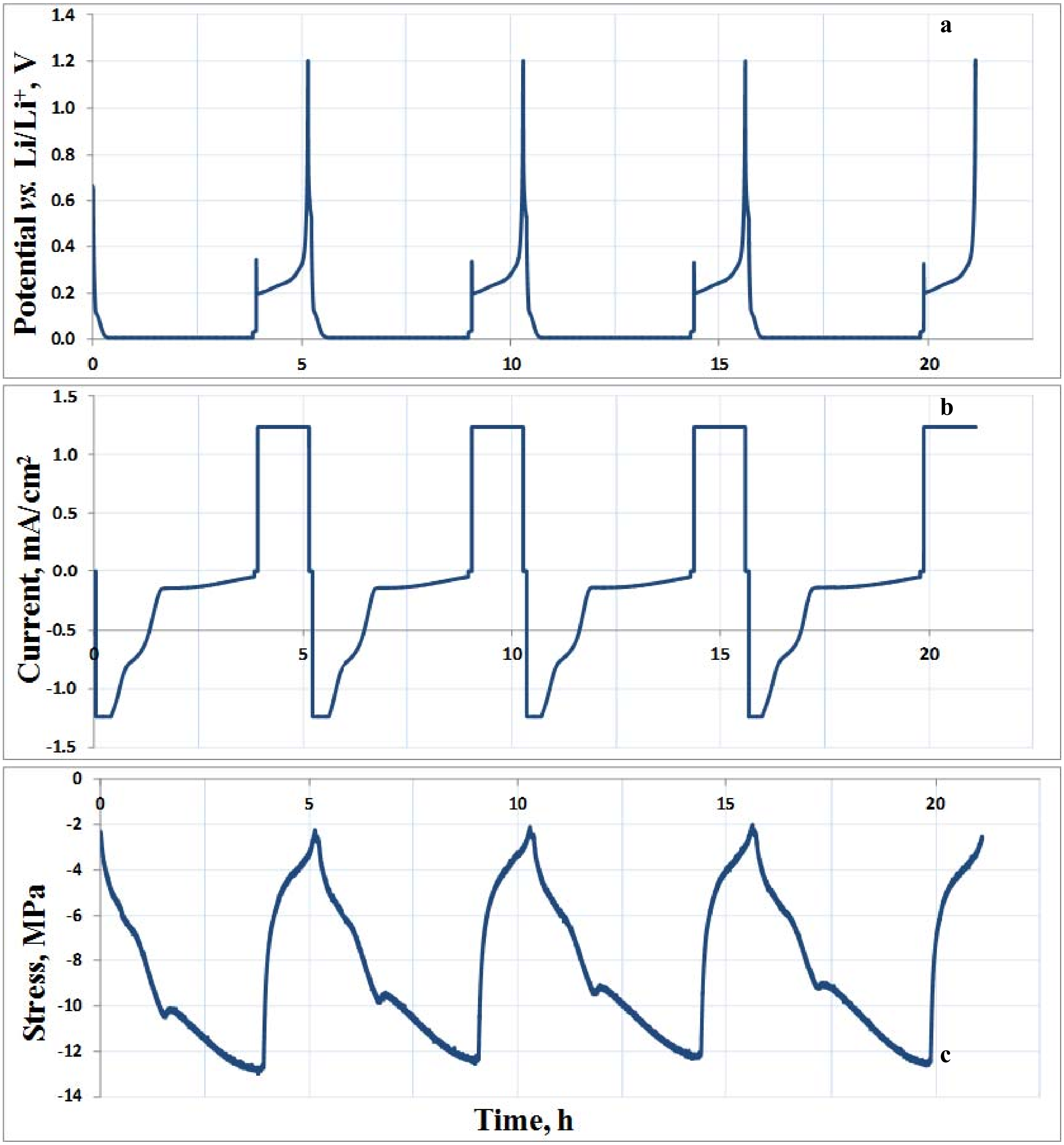

**Figure 8**

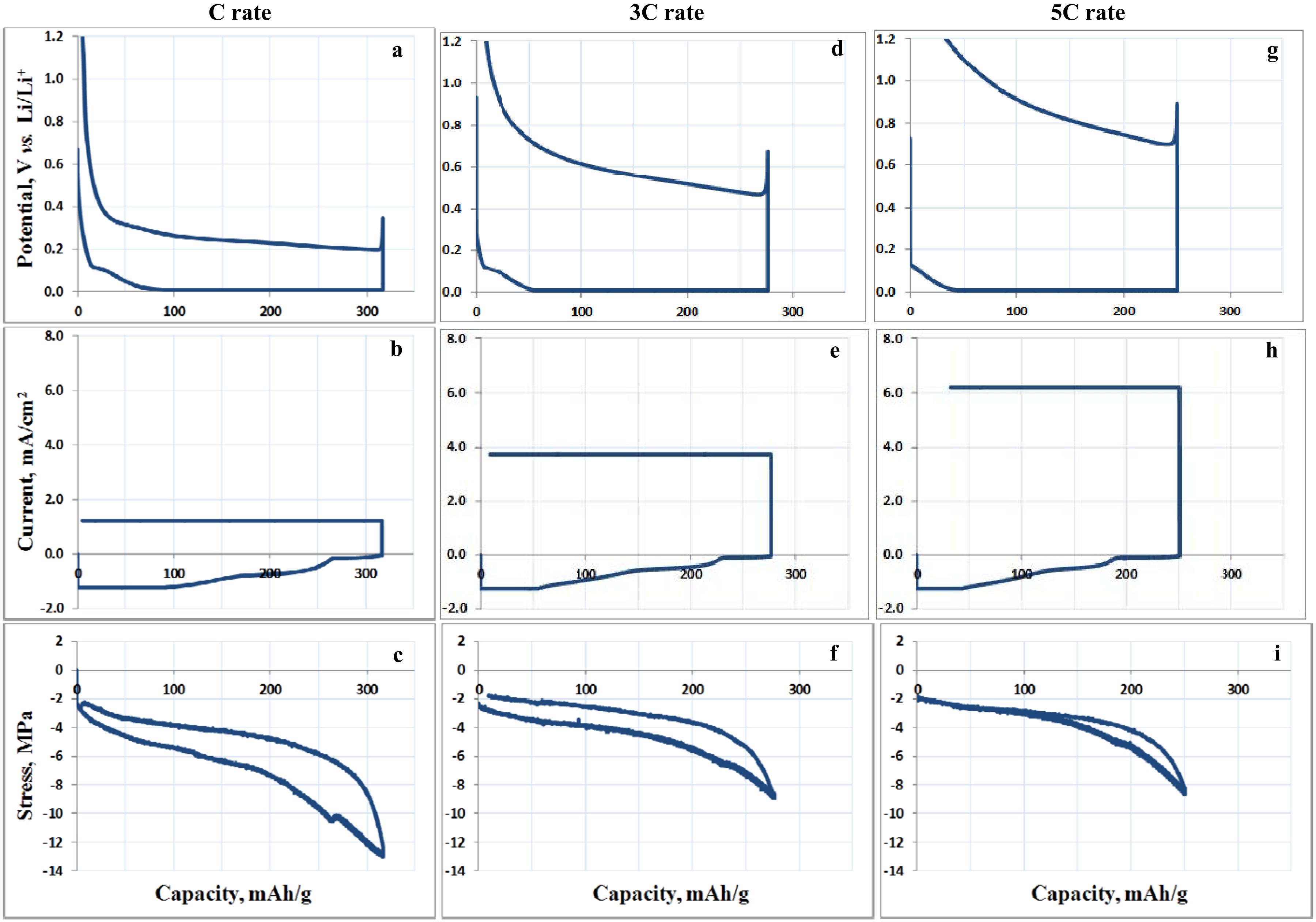